\documentstyle[aps,epsfig]{revtex}

\title{New scenario for  high-$T_c$ cuprates :
electronic topological  transition as a motor for
anomalies in the underdoped regime.}

\author{F. Onufrieva , P. Pfeuty  \and M. Kisselev}
\address{Laboratoire Leon Brillouin CE-Saclay 91191
 Gif-sur-Yvette France}

\begin{document}

\twocolumn[\hsize\textwidth\columnwidth\hsize\csname@twocolumnfalse\endcsname
\maketitle


\begin{abstract}
We have discovered a new nontrivial aspect of  electronic
topological transition (ETT) in a 2D free fermion system on a
square lattice. The corresponding exotic quantum critical
point, $\delta$=$\delta_c$, $T$=$0$,
($n$=$1-\delta$ is an electron concentration)
 is at the origin of
  anomalous behaviour in the interacting system
on one side of ETT, $\delta<\delta_c$. The most important
 is an appearance of the line
of characteristic temperatures, $T^*(\delta) \propto
\delta_c-\delta$. Application of the theory to  high-$T_c$ cuprates reveals
a striking similarity to the observed experimentally
behaviour in the underdoped regime (NMR and ARPES).
\end{abstract}

\pacs{74.25.-q, 74.72.-h, 74.25.Dw, 74.25.Ha}
]



\date{\today}

This is a particularly exciting time  for high-$T_c$.
The experimental knowledge  converges. Almost all
  experiments, NMR \cite{NMR,Takigawa}, ARPES
 \cite{ARPES}, infrared conductivity
\cite{conductivity} etc.
provide an evidence for the existence of a
characteristic energy scale  $T^*(\delta)$ in
the underdoped regime  ($\delta$ is  hole
doping).
Below and around the
line $T^*(\delta)$ the "normal" state (i.e.
above $T_c$) has properties fundamentally incompatible
with the present understanding of metal physics.
The field has reached the point when a consistent theory
is necessary to understand this exotic from  theoretical
point of view but quite well defined  from
experimental point metallic behaviour.
The issue has a significance beyond the field of
high-$T_c$ superconductivity - the fundamental question
arises : what kind of metallic behaviour is there,
in addition to the well understood Fermi liquid?

In the paper we propose our variant of the answer.
We reexamine a free electron 2D system on a square
lattice
 with hopping beyond nearest neighbors.
 We show that when varying the electron concentration
 defined as $1-\delta$, the system undergoes   an 
 electronic topological
transition (ETT)  \cite{Lifshitz} at a critical value $\delta$=$\delta_c$. The
corresponding $T$=$0$ quantum critical
point (QCP)  combines several  aspects
of criticality. The first standard one is related to
 singularities in thermodynamic
properties, in  density of states at $\omega$=$0$
(Van Hove singularity),
to  additional singularity in the
  superconducting
(SC) response function (RF) \cite{singularity}.
 The second nontrivial aspect
 is: the same QCP {\bf is the  end
of the critical line} $T$=$0$, $\delta>\delta_c$ each point
$\delta$ of which is characterized by    static
Kohn singularity (KS) in polarizability of 2D free
fermions.
[What we mean as a static KS is a  singularity
at the wavevector
  connecting two points of Fermi surface (FS) with
parallel tangents \cite{Kohn}].
 The two aspects of criticality are not related with each other.
It is the latter aspect (never considered before) which as we will show
is a motor for anomalous behaviour in the regime $0<\delta<\delta_c$
of the system of noninteracting
and  interacting electrons (or of any fermion-like quasiparticles 
e.g. of those \cite{Onufrieva}
appearing in
the $t-t'-J$
model 
describing the strongly correlated $CuO_2$
plane responsible for main physics in the cuprates).
The found anomalies
 have a striking similarity
to  anomalies in the underdoped
high-$T_c$ cuprates. The effect  exists
in all cases $t' \neq 0$ or/and $t'' \neq 0$ ... except
for  special sets of the parameters corresponding to the perfect
nesting in FS (including
 $t'$=$t''$=...$\rightarrow 0$) studied in many papers, see e.g. 
\cite{Ruvalds}. For such sets
 the QCP loses the latter aspect of criticality
 and the anomalies  disappear.

A starting point is a 2D electron system
on a square lattice
 with hopping beyond nearest neighbors

\begin{equation}
\epsilon_{{\bf k}} = \ -2t (\cos k_x + \cos k_y ) - 4t' \cos k_x
\cos k_y -...\ \label{1}
\end{equation}
 For any set of the parameters $t$,
$t'$, $t''$,...
  the dispersion law is characterized
 by two different saddle points (SP's) located at
 $ ( \pm\ \pi, 0)$ and $(0, \pm  \pi)$
   with the energy $\epsilon_{s}$.
When we vary the chemical potential $\mu$ or the energy
 distance from
the SP, $Z=\mu-\epsilon_{s}$,
 the topology
of the FS changes when $Z$ goes from $Z>0$ to $Z<0$ through
the critical value $Z=0$. In
vicinities of  SP's the dispersion law is  :

\begin{equation}
\tilde{\epsilon}({\bf k}) =\epsilon_{{\bf k}}-\mu= -Z +a k_\alpha^2 -
b k_\beta^2,
\label{2}
\end{equation}
where $\bf k$ is measured from  $(0,\pi)$
($\alpha=x$, $\beta=y$) or from $(\pi,0)$
($\alpha=y$, $\beta=x$).
 Explicit expressions for $a$ and $b$ depend on $t$, $t'$,...
We consider the general case : $a \neq 0$, $b \neq 0$, $a \neq b$. We choose $a>b$  corresponding to $t'/t<0$.

The $T=0$ ETT  has  several characteristic aspects.
The first (trivial) one
 is related to the {\bf local change of FS topology}
in the vicinity of SP.
 This leads to
 divergences in  thermodynamic
properties,
in  density of states at $\omega=0$ etc.
 From this point of view  the corresponding
QCP is of a gaussian type with the dynamic exponent $z=2$.

The nontrivial aspect
 is related to  {\bf mutual change} in  topology of FS in
vicinities of two different SP's and reveals
itself when considering the electron polarizalibility

\begin{equation}
\chi^0({\bf q},\omega)=\frac{1}{N}\sum_{{\bf k}}\frac{n^{F}(\tilde{\epsilon}_{{\bf k}})
-n^{F}(\tilde{\epsilon}_{{\bf {q+k}}})}{\tilde{\epsilon}_{{\bf q+k}
}-
\tilde{\epsilon}_{{\bf {k}}}-\omega-i0^+}.
\label{3}
\end{equation}
We show that the latter
 has a  square-root  singularity
at $\omega$=$0$  and  
wavevector  $\bf q$=${\bf q}_m$ in a
 vicinity of ${\bf Q}$=$(\pi,\pi)$ {\bf for any $Z$ on the semiaxis $Z<0$}:
$\chi^0({\bf q}_{},0)-\chi^0({\bf q}_{m},0)\propto
 \sqrt{|{\bf q}_{m}-{\bf q}|}$ for
${\bf q}<{\bf q}_{m}$.
It is a  static  KS in the 2D electron system.
The locus of
the wavevectors ${\bf q}_m$ in the BZ
is
 a closed curve around
${\bf Q}$ with
$|{\bf Q}-{\bf q}_m|$$\propto \sqrt{\mid Z \mid}$. With
decreasing $|Z|$ the  closed curve shrinks
and is reduced to the point ${\bf q= Q}$ at $Z=0$ where
 $\chi^0({\bf q},0)$ diverges
 logarithmically.
The  curve of the static  KS's with $\bf q$ close
to $\bf Q$ does not reappear  for $Z>0$:
 $\chi^{0}({\bf q},0)$  is peaked at  $\bf q$=$\bf Q$ in
an intimate vicinity of ETT and  it exhibits a wide plateau around
${\bf q=Q}$ for larger $Z$.
To illustrate this we show in Fig.1 the  $\bf q$
 dependence of $\chi^{0}({\bf q},0)$ calculated
 based on (\ref{3}) and  (\ref{1}).
[We use the model with
only  $t'$$\neq 0$ being a generic model for
 the  family: $a$$\neq 0$,  $b$$\neq 0$,  $a$$\neq b$.]
The discussed above curve is the closest to
${\bf q=Q}$ curve of singularities in Fig.1a.
 In the plots  one sees
 only a quarter of the
picture around ${\bf q= Q}$; to see the {\bf closed}
 curve  around
$(\pi,\pi)$ one has to consider the extended BZ.
 [Few
other curves of  KS's seen in Fig.1
 are not sensitive to ETT,
 we  discuss them elsewhere.] 

\vspace*{-15 mm}
\begin{figure}
\begin{center}
\epsfig{%
file=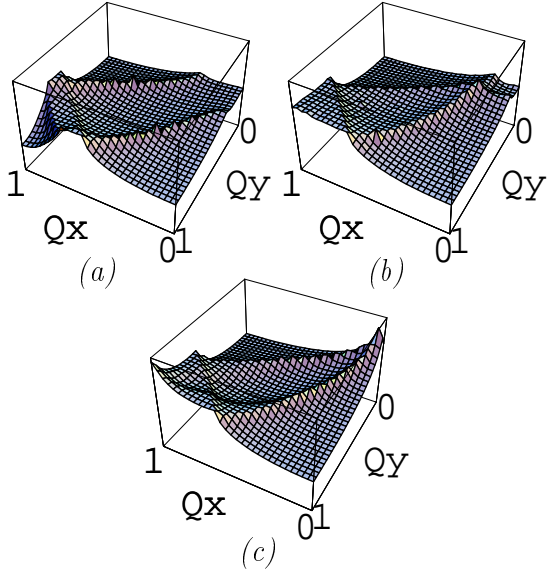,%
figure=nfig1.ps,%
height=14.5cm,%
width=9.5cm,%
angle=0,%
}
\\
\end{center}
\vskip -6.2 cm \caption{${\bf q}$ dependence of $\chi^{0}({\bf
q},0)$ through the  Brillouin zone for (a)$Z<$$0$,
 (b)$Z>$$0$, (c)$Z$=$0$.
 $Q_x$=$q_x/\pi$,  $Q_y$=$q_y/\pi$. The point
${\bf q=Q}$ corresponds to the left corner. ($t'/t=-0.3$)}
\label{f1}
\end{figure}

As a result, the point $Z=0$,
$T=0$ turns out to be the end point of the critical line
$Z<0$, $T=0$.

Paradoxically, the {\bf absence} of the discussed curve of static KS's for
$Z>0$ leads to  anomalous behaviour of the system on this side
of QCP. To see this let's calculate
 $\omega$ dependencies of
 $Re\chi^0({\bf q},\omega),
Im\chi^0({\bf q},\omega)$ and
$C(\omega)$=$Im\chi^0({\bf q},\omega)/\omega$  for the characteristic
for this regime wavevector ${\bf q}$=${\bf Q}$.
 Results are
shown in Fig.2a. One can
see that all functions are singular at some energy $\omega_c$.

Calculations  with the hyperbolic
spectrum  (\ref{2}) give the following  expression:
$Im\chi^{0}({\bf Q},\omega)=
 F(\omega/\omega_c,b/a)/2\pi t$,
 $Re\chi^0({\bf Q},\omega)=Re\chi^0({\bf Q},\omega_c)
-\Phi(\omega/\omega_c,b/a)/t$ with
$$
F(x,y) =
 \left\{
\begin{array}{cc}
\ln{\frac{\sqrt{1+xy}+\sqrt{1+x}}{\sqrt{1-xy}+\sqrt{1-x}}},
&  0 \leq x \leq 1\\
\ln{\frac{\sqrt{1+xy}+\sqrt{1+x}}{\sqrt{x(1-y)}}}, &  x \geq 1
\end{array}
\right.
$$
\begin{equation}
\Phi(x,y) =
 \left\{
\begin{array}{cc}
\gamma_1(y)(1-x^2), &  x -1 <0\\
\gamma_2(y)\sqrt{x-1},
&  0 \leq x-1 \ll 1
\end{array}
\right.
\label{4}
\end{equation}
($\gamma_{1,2}(y) \ll 1$). The {\bf new energy scale} which appears and corresponds to the
singularities in Fig.2a is given by

$$
 \omega_c=Z(1+b/a).
$$

The singularities at $\omega= \omega_c$ are  dynamic 2D KS's.

\begin{figure}
\epsfig{%
file=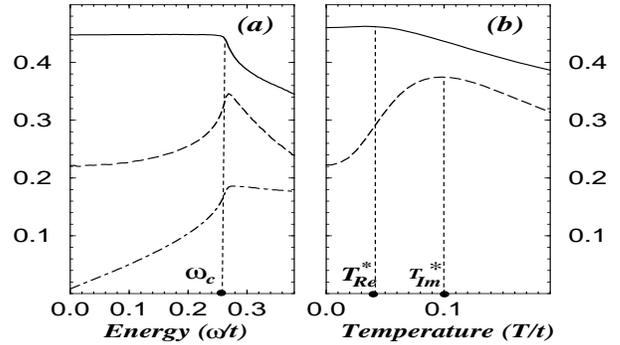,%
figure=nfig2.eps,%
height=4.5 cm,%
width=8 cm,%
angle=0,%
}
\\
\vspace*{-2mm}
\caption{$Re\chi^0({\bf Q},\omega)$ (solid),
 $Im\chi^0({\bf Q},\omega)$ (dot-dashed) and
 $C(\omega)$=$Im\chi^0({\bf Q},\omega)/\omega$ (dashed)
 in the regime $Z>0$
 (a) as a function of $\omega$ for $T=0$,
 (b) as a function of $T$ for $\omega \rightarrow 0$
($Im\chi^0({\bf q},0)=0$).
 Here $t'/t=-0.3$,
$Z/t=0.21$.}
\label{f2}
\end{figure}

The dynamic KS's at $T=0$ are transforming into
static Kohn anomalies at finite temperatures which again
scale with $Z$, see Fig.2b.
 When comparing with Fig.2a
one can see that the behaviour is similar
being  smoothed by the effect of finite $T$
(both behaviour are very different from the standard one
for noninteracting system).
 The important difference is
that the characteristic temperatures
of the Kohn anomalies for
$Re\chi^0({\bf Q},0)$ and for
$\lim_{\omega \rightarrow 0}Im\chi^0({\bf Q},\omega)/\omega$
being both scaled with $Z$
$$
T^*_{Re} = A Z, \hskip 0.5 cm T^*_{Im} = B Z, \hskip 0.5 cm A<B
$$
are different on the contrary to the characteristic energy
of the KS's
at $T$=$0$ (that is an usual effect of finite $T$).

 Another remarkable signature of {\bf asymmetry in $Z$} is
the following. Taken for the characteristic for
each regime  wavevector, ${\bf q}$=${\bf q}_m$ for $Z$$<0$ and
${\bf q=Q}$ for $Z>0$,
  $\chi^0({\bf q},0)$
decreases rapidly with $|Z|$ for $Z$$<0$ (ordinary behaviour) while
for $Z$$>0$ it  remains {\bf
practically constant} (and quite high) for $Z$ not too small.
Moreover   for finite $T$, $\chi^0({\bf Q},0)$
has a maximum at $Z$=$Z^*(T)$$>0$
(that is very unusual). As a result of the described $T$ and $Z$
dependencies of $\chi^0({\bf Q},0)$ in the regime $Z>0$, the lines
$\chi^0({\bf Q},0)$=const have an  unusual form in the  $T-Z$
plane: they develop rather around the "critical" lines
$T_{Re}^*(Z)$ and $T_{Im}^*(Z)$ than around the QCP, $T$=$0$, $Z$=$0$.

On the contrary, a behaviour of SC RF (in both cases isotropic s-wave
or d-wave symmetry) is symmetrical in $Z$ being related to the first
aspect of ETT. For the same reason the SC RF decreases quite
rapidly with
increasing a distance from QCP, i.e. with increasing $T$ and $|Z|$.

Above we considered a system of noninteracting electrons. In fact
the same picture takes place for any system of fermion or
fermion-like quasiparticles which dispersion law is determined by
the topology of 2D square lattice and has a form (\ref{1}). In
\cite{Onufrieva} where we discuss some problems of strongly
correlated systems we show that such quasiparticles (with spin and
charge) do exist in the $t-t'-J$ model describing the strongly
correlated 
  $CuO_2$ plane.
 On the other hand,
the  shape of FS observed by ARPES  does imply the existence of
nnn hopping $t'\neq 0$,  so that the condition of the asymmetry $a
\neq b$ necessary for the existence of the discussed ETT is
fulfilled. Moreover this shape  implies $t'/t<0$, the case  for
which the  critical doping $\delta_c$ is positive. Below we will
 pass from the energy distance from ETT, $Z$, to the
doping distance, $\delta_c-\delta$, using large FS
condition:
$1-\delta$=$2\sum_{\bf k}n^F(\tilde{\epsilon}_{\bf k})$
\cite{Onufrieva}.

\vspace*{-5mm}
\begin{figure}
\epsfig{%
file=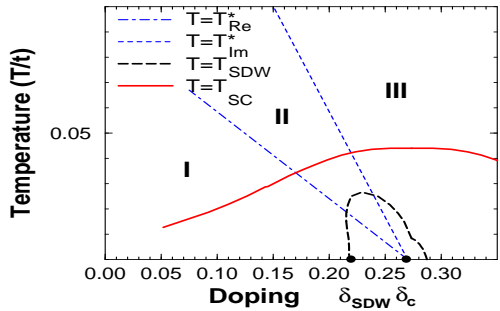,%
figure=nfig3.eps,%
height=4.5cm,%
width=7.cm,%
angle=0,%
}
\\
\vspace*{-3mm}
\caption{Phase diagram
with the lines of SDW
and d-wave SC instabilities
 and the  lines  $T^*_{Re}(\delta)$,
  $T^*_{Im}(\delta)$ ($t'/t=-0.3$, $t/J=1.9$).
  We consider only the
metallic part of phase diagram (for discussion about a
passage from AF  localized-spin state at low doping to the metallic
state with large FS for intermediate doping see [8]).}
\label{f4}
\end{figure}

Let's consider now the system in the presence of
interaction.   A quite trivial consequence of
the ETT is a developing of density wave (DW) and SC instabilities around
the point $\delta$=$\delta_c$, $T$=$0$. [The effects
are related to the logarithmic divergence of $\chi^0({\bf Q},0)$
and $\ln{Z} \ln{T}$
divergence of the SC RF as
 $T \rightarrow 0$, $Z \rightarrow 0$.]
Nontrivial consequences concerning the DW degrees of freedom and
related to the Kohn singularity aspect of ETT
 are:
{\it (i) strong asymmetry between regimes $\delta<\delta_c$ and
$\delta>\delta_c$ and (ii)
  very long (in doping and temperature) memory about
DW instability in the disordered state on one side of ETT,
$\delta<\delta_c$}. To see this let's consider the electron-hole
RF which in
the simplest  RPA
  approximation   is given by
$\chi({\bf q},\omega)=
\chi^0({\bf q}, \omega)/(1+V_{\bf q}\chi^0({\bf q},
 \omega))$.
In the case of interaction $V_{\bf q}$ in a triplet (singlet)
channel the instability and normal state fluctuations
are of SDW (CDW) type. We will consider the former
interaction: $V_{\bf q}$=$J_{\bf q}$=2$J(\cos q_x +\cos q_y)$ ($J>0$)
 as strongly supported
 by neutron scattering
and NMR experiments for the cuprates and on the other hand as
an interaction  between the discussed
above quasiparticles
in the $t-t'-J$ model \cite{Onufrieva}. For such interaction
both instabilities d-wave SC  (see details in \cite{Onufrieva})
and SDW  take place around QCP. Due to the symmetry of
SC RF in $Z$, $T_{sc}(\delta)$ is symmetrical on two sides
of $\delta_c$ with a maximum  at $\delta=\delta_c$, see Fig.3.
Therefore the regimes $\delta<\delta_c$ and $\delta>\delta_c$
can be considered
as underdoped and overdoped, respectively.
 On the contrary,  the
line of  SDW
 instability,
$T_{SDW}(\delta)$,   given by
$\chi^0({\bf q}, 0)=-1/J_{\bf q}$ (${\bf q=Q}$ for
$\delta<\delta_c$ and ${\bf q=q}_m$ for
$\delta>\delta_c$)
 has an
 anomalous form in the regime $\delta<\delta_c$: it
develops rather around the lines $T_{Re}^*(\delta)$
and $T_{Im}^*(\delta)$ than around QCP (see Fig.3)
 reproducing the form of  lines
 $\chi^0({\bf Q}, 0)=const$
discussed above.

When at certain doping, $\delta$=$\delta_{SDW}$,
 the ordered SDW solution
disappears, it is the  disordered metallic state which keeps this
type of behaviour:
the regime $T_{Re}^*(\delta)<T<T^*_{Im}(\delta)$ (II)
turns out to be a regime of a {\bf minimum disorder}  and the regime $T<T_{Re}^*(\delta)$ (I) a regime of
{\bf a reentrant in temperature  quantum SDW liquid}.
 Indeed, two most important parameters
characterizing SDW liquid:  $\kappa^2=1-|J_{\bf Q}|
 \chi^0({\bf Q},0)$
 describing a "proximity"
 to the  SDW
instability and $\Gamma_{{\bf Q}}=\kappa^2/C(0)$
describing
a relaxation energy behave in a reentrant way in increasing $T$:
$\kappa^2$ decreases (slightly)  with $T$ until
 $T_{Re}^*(\delta)$ and $\Gamma_{{\bf Q}}$ decreases
(strongly)
 until
 $T_{Re}^*(\delta)<T^*_\Gamma<T^*_{Im}(\delta)$  as if the system would
move towards an ordered phase.
However, it does not reach it, the reentrancy
stops and the system passes to the regime II
of a minimum disorder above which a standard
 disordered state behaviour is restored (regime III).
On the other hand, the quantum SDW liquid state
in the regime I is practically {\bf frozen in doping}
due to the very weak dependence of $\kappa^2$ on doping.
 As a result the  disordered metal state in the
regime  $\delta<\delta_c$
keeps
a strong memory about the ordered SDW phase (and therefore
develops  strong critical SDW fluctuations) very far  in doping
and in temperature. On the contrary, in the regime $\delta>\delta_c$
 the memory about SDW
instability and the corresponding fluctuations
disappear rapidly
due to the sharp decrease of
 $\chi^0({\bf q},0)$ with increasing $\delta-\delta_c$ and $T$.
The same is valid in both regimes $\delta>\delta_c$
and  $\delta<\delta_c$ for SC fluctuations due to the
discussed in the first part behaviour of SC RF as
a function of $T$ and $|Z|$.
Therefore, although the SDW phase itself is energetically
unfavorable with respect to the SC phase
(except of the case of very high $J/t$), the metal state
above  $T_{sc}$ in the underdoped regime
is a precursor of the SDW phase rather than
of the SC phase.

\vspace*{-2mm}
\begin{figure}
\epsfig{%
file=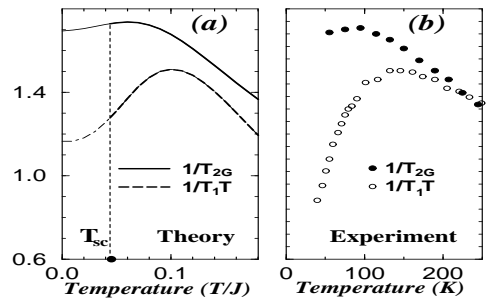,%
figure=nfig4.eps,%
height=4.cm,%
width=7cm,%
angle=0,%
}
\\
\vspace*{-2mm}
\caption{$1/T_1T$ and $1/T_{2G}$ (a) calculated for $\delta=0.15$
($t'/t=-0.3$, $t/J=1.9$) (should be considered
only above $T_{sc}$)
and  (b) taken from NMR for $YBCO_{6.6}$ [2].}
\end{figure}

The lines $T^*_{Re}(\delta)$ and $T^*_{Im}(\delta)$
are basic lines for anomalies in the disordered
metallic state.
To demonstrate how the anomalies appear for different
properties
we consider some examples. In Fig.4a we show calculated
quasistatic magnetic characteristics
 corresponding to the
 measured  by NMR $1/T_1T$ and $1/T_{2G}$
on copper as  functions of $T$. A physical reason for
a slight increase of $1/T_{2G}$
 extending until $\approx T^*_{Re}$ and a much stronger increase of
$1/T_1T$ extending until $T \approx T^*_{\Gamma}$ is the discussed above
reentrant behaviour of   $\kappa^2$ and $\Gamma_{{\bf Q}}$ with $T$.
The theoretical behaviour
is very close to that observed experimentally (Fig.4b)
 and explains it in fact for the first time.

In Fig.5 we show an  electron spectrum
calculated for the ordered SDW phase (a)
and for the disordered metal state (namely for
the regime II) (b).
 For the  ordered phase the spectrum  is given by:
$\varepsilon_{1,2}=(\epsilon_A+\epsilon_B)/2
\pm
\sqrt{(\epsilon_A-\epsilon_B)/2)^2+
\Delta^2}$
($\epsilon_A({\bf k}) \equiv \epsilon({\bf k})$,
$\epsilon_B({\bf k}) \equiv \epsilon({\bf k+Q})$)
with the gap $\Delta$  determined selfconsistently
in the usual way. For the disordered state the
"spectrum" is obtained from maxima of electron spectral
functions  strongly renormalized
due to  interaction with the described above SDW
fluctuations.
The  characteristic form of the spectrum in both cases
is a result of a hybridization of two parts of the bare spectrum
in the vicinity of two different SP's $(0,\pi)$ and $(\pi,0)$.
The hybridization is static for the ordered SDW phase
and is dynamic for the disordered state.
[Details about the pseudogap opening in the
disordered state
and its behaviour with $T$ and $\delta$  will
be a subject of a separate paper.]
 The spectrum is in  excellent agreement with
ARPES data, see Fig.5c (ARPES measures only the part corresponding to
$\omega<0$). The effect of splitting into two branches
and of the pseudogap disappears quite rapidly
in the regime $\delta>\delta_c$ due
to the rapid  weakening of  SDW fluctuations. Due to the same reason
it disappears roughly above $T^*_{Im}(\delta)$. Both facts  agree
with experiments for the cuprates.

\begin{figure}
\hspace*{-1.7cm}
\epsfig{%
file=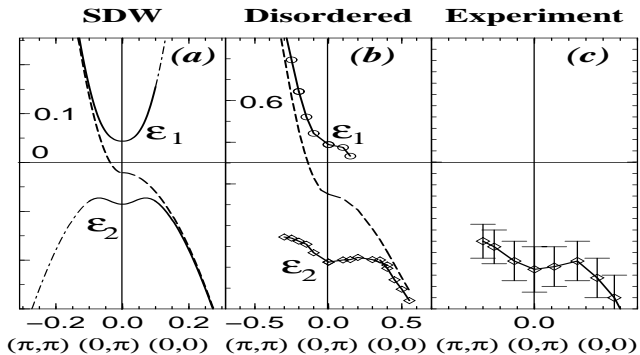,%
figure=nfig5d.eps,%
height=45mm,%
width=9.9cm,%
angle=-0,%
}
\\
\vspace*{-4mm}
\caption{Electron spectrum $\varepsilon({\bf k})/t$ along $\Gamma - X$ symmetry lines,
 (a) in SDW phase ($Z/t$=$0.03$ ($\delta$=$0.25$), $T$=$0$), (b) in
the
metallic state above $T_{sc}$ ($Z/t$=$0.3$ ($\delta$=$0.1$), $T/t$=$0.15$),
(c) ARPES data [10] for underdoped $Bi_2Sr_2CaCu_2O_{8+delta}$ above $T_{sc}$.
The   dashed lines correspond to
 the bare spectrum, the thin line  in (a)
 to the spectrum with
the spectral weight less than $0.1$. $t'/t$=$-0.3$, $t/J$=$1.8$,
wavevectors are taken in unit of $\pi$.} \label{f5}
\end{figure}

Let's discuss a behaviour of $Im\chi({\bf q},\omega)$, the
characteristics measured by inelastic neutron scattering (INS).
As follows from the previous analysis,
below  $T^*_{Im}$ it has a maximum at $\omega$=$\omega_0$$\propto$$\kappa^2$
(being peaked at ${\bf q}$=${\bf Q}$). So far as $\kappa^2$
almost does not change with $\delta$, the position
of the peak does not as well. This agrees
with INS data
and  explains (for the first
time) the existence of the characteristic energy ($\sim 30$ meV)
above $T_{sc}$ for all $\delta$, see e.g. the
summarizing picture in Fig.25 in \cite{INS}.
As was emphasized before, strong SDW fluctuations
 disappear in the
overdoped regime $\delta>\delta_c$. In the underdoped regime
they disappear (or strongly diminish) above $T^*_{Im}(\delta)$.
Both facts are in
a good agreement with INS.

Summarizing, the simple picture arising from the effect of
ETT in 2D electron system on a square lattice gives
 a unified vision
of  normal state anomalies in the underdoped high-$T_c$ cuprates for both
magnetic and electronic properties. We succeed to explain the
temperature anomalies in  $1/T_1T$ and $1/T_{2G}$ NMR
characteristics, some crucial features of INS in the normal state,
the disappearance of magnetic fluctuations in the overdoped
regime, an opening of a pseudogap in electron spectrum, the shape of the
latter in a vicinity of $(0,\pi)$, the disappearance
 of the pseudogap   in the overdoped
regime. All these are most nontrivial experimental results.
Regarding that the theory does not use any 
external  phenomenological hypothesis and only two
microscopical parameters $t'/t$ and  $t/J$, the similarity
between the theoretical results and experiments seems  quite
remarkable. We emphasize that the effect exists
for any $t'/t$, $t''/t$,... except of two limit cases: (i)
isotropic one $a$=$b$  in eq.(2)
 ($t'$=$t''$=...=0) and (ii) extreme anisotropic one
 $a$=$0$ or $b$=$0$. Although ETT exists in both cases,
the corresponding QCP's belong to different classes of universality.
For $a$=$b$ (nesting) the behaviour is symmetrical in $Z$, the
anomalous regime discussed in the paper disappears.

\end{document}